\documentclass{iopart}

\usepackage{graphicx}
\usepackage{url}
\usepackage{array}
\begin{document}

\title[A giant red shift and enhancement of the light confinement in
a dielectric array] {A giant red shift and enhancement of the light
confinement in a planar array of dielectric bars}

\author{Vyacheslav V Khardikov$^{1,2}$, Ekaterina O Iarko$^1$ and Sergey L Prosvirnin$^{1,2}$}

\address{$^1$ Institute of Radio Astronomy of National Academy of Sciences of Ukraine, Kharkiv 61002, Ukraine}
\address{$^2$ Karazin Kharkiv National University, Kharkiv 61077, Ukraine}
\ead{khardikov@univer.kharkov.ua, yarkokatya@rian.kharkov.ua and
prosvirn@rian.kharkov.ua}

\begin{abstract}
The results of the research of resonant phenomena in double-periodic
subwavelength planar structures made up of paired dielectric bars
are presented. For the first time the existence of high Q-factor
trapped mode resonances is revealed in these low-loss entirely
dielectric structures. A great red shift of the trapped mode
resonance of the structure is observed as compared with the resonant
wavelength of the periodic structure with only one dielectric bar
per unit cell. This shift of the resonant wavelength is caused by a
strong coupling of the electromagnetic fields in the adjacent
dielectric-bar resonators.
\end{abstract}
 \maketitle

\section{Introduction}

Thanks to a striking progress in nanotechnology, the optically thin
layers of materials can be structured with periodic pattern in order
to design planar metamaterials. The planar metamaterials (also known
as metafilms) are an impressive modern object, which is driven by
certain fascinating facilities such as, e.g., the anomalous
reflection and refraction \cite{capasso-2011-lpw}, the unusual
asymmetric transmission \cite{fedotov-2006-asymmetry,
fedotov-2007-asym-plasmon}, and light reflection that does not
change the phase of the incident wave \cite{schwanecke-2007-mirror}.

Typically, a planar metamaterial assigned for visible and near
infrared wavelengths is a plasmonic structure designed on the basis
of a periodic array of complex-shaped resonant nanowire metallic
particles. The main factor responsible for the spectacular
properties of metafilms is some resonant interaction of light with
the patterned layer. Moreover, numerous envisioned applications of
planar metamaterials do require the high Q-factor resonances and a
strong confinement of intensive electromagnetic fields. First of all
they concern both the projects of gaining or lasing devices such as
the spaser \cite{bergman-2003-spaser, zheludev-2008-lspaser} and the
devices characterized by the bistable reflection and transmission in
order to control light with light \cite{tuz-2010-obi,
kivshar-2010-osi}, which could be realized by incorporating some
active medium or nonlinear inclusions in periodic structures of
resonant metafilms. Another domain, aging a decade, concerns the
extremely sensitive chemical and biochemical sensors
\cite{flory-2011-opo}.

However, losses are orders of magnitude too large for the envisioned
applications. Typically, the Q-factor of the resonances excited in
plasmonic structures is small because of high radiation losses and
huge energy dissipation inherent to metals in the visible and near
infrared wavelength ranges.

The usual resonance field enhancement inside a planar metamaterial
may be extremely increased by involving structures which bear the
so-called trapped modes \cite{stockman-2001-lvd, yariv-2009-git}.
The excitation of the trapped mode resonances in planar
double-periodic structures with broken symmetry was found both
theoretically \cite{prosvirnin-01-mla, prosvirnin-2003-rcm,
arnaut-2004-hps} and experimentally \cite{fedotov-2007-stm} in
microwaves. In particular, those typical peak-and-trough Fano
spectral profile resonances can be excited in the periodic array
composed of twice asymmetrically-split metal rings. Their specific
character arises from some destructive interference of the radiation
by the anti-phased currents in metallic elements of a subwavelength
periodic cell.

In the fully symmetric structure, the coupling of the trapped mode
currents can be infinitesimal for a wave incident from free space.
Thus, the Q-factor of the resonance is limited only by the
dissipative losses of the structure. This unique property leads to
the analogy between the trapped mode resonance and the metastable
energy level in atomic systems, manifesting an electromagnetically
induced transparency. Recently, some effective media exhibiting the
EIT-like properties and the desired "light slowing" metamaterials
that use the trapped mode plasmonic arrays, have been proposed
\cite{zhang-2008-pit, papasimakis-2008-mao}.

Now, the existence conditions and the spectral properties of trapped
mode resonances are investigated in detail in the suitably
structured planar plasmonic metamaterials developed for the near-IR
range \cite{khardikov-2010-radiophysics, zhang-2010-double_bars,
khardikov-2010-jop}. The Q-factor of the trapped mode resonance
essentially exceeds that for the ordinary resonance, but its value
is not greater than several tens because of dissipative losses.


Since the intrinsic dissipative losses in plasmonic structures are
unavoidable, the idea of their compensation using certain parametric
processes and gain media was proposed \cite{shalaev-2006-cli,
ziolkowski-2007-tda}. Experimental demonstrations of the
compensation of the absorption losses in a trapped-mode metallic
metamaterial by using optically pumped semiconductor quantum dots
were presented in \cite{plum-2009-tls, tanaka-2010-meo}. The
observed narrowing of the quantum dots photoluminescence spectrum
evidences of the Q-factor increase in the pumped structure.

Fortunately, the use of plasmonic structures is not a single
possible way to develop the thin planar metamaterials to ensure a
strong enhancement of the confined resonant field. One evident way
to produce low-dissipative structures, with similar electromagnetic
properties is to use entirely dielectric elements in designing the
double-periodic structures maintaining the trapped mode resonances.
It was shown recently, by simulation, that the microwave left-handed
media may be constructed on the basis of cubic high dielectric
resonators \cite{kim-2007-som} or dielectric rings and rods put
together \cite{marques-2010-art-magnetism} instead of the classical
split-ring metallic resonators.

This paper is aimed at involving in the family of the high-Q trapped
mode planar metamaterials the new and highly desirable low-loss
structures made up of entirely dielectric elements.

The resonances in the plasmonic and dielectric structures have a
considerably different nature. A plasmon-polariton excitation
propagating along metallic surfaces replicates their shape.
Therefore, the complex-shaped metallic elements can be used to
provide a resonant interaction light with a periodic structure in a
deeply subwavelength range. On the contrary, complications of the
shape of dielectric elements do not involve a substantial increase
of the resonant wavelength, like in the case of metallic elements.

The paper studies the excitation conditions and the main properties
of the near-infrared trapped-mode resonance in a low-loss
subwavelength planar array with the unit cell that includes a pair
of resonant dielectric bars.

\section{The problem statement and the method of solution}

\begin{figure}
\centering
\includegraphics[width=4.0in]{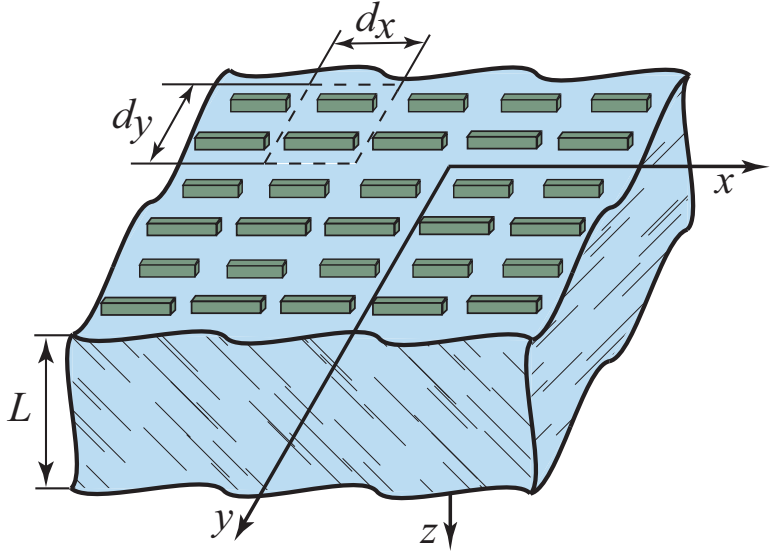}
\caption{Geometry of double-periodic planar structure with two
dielectric bars in the periodic cell.} \label{fig1}
\end{figure}

Let us consider a dielectric bar of square cross section as a
constituent element of the entirely dielectric array. Thus the
problem of light diffraction by a double-periodic planar array of
dielectric bars placed on a substrate with thickness $L$ is under
investigation (see Fig.~\ref{fig1}). The unit cell of the array
includes a pair of dielectric bars which have different in length
but identical in cross-section and material. The longer bar length
is $h_1$ and the shorter $h_2$. The sizes of the square periodic
cell are $d_x=d_y=975$~nm. The distance between the longer and the
shorter bars of the pair is 195~nm for all of the two-element
dielectric arrays studied below. The periodic cell is symmetric
relative the line drawn through the cell center parallel to the
$y$-axis. The normal incidence of a linearly $x$-polarized plane
wave is considered. The resonant response of the array is studied in
the near infrared wavelength range from 1000~nm to 3000~nm. The
substrate material is assumed to be synthetic fused silica. Its
refractive index is approximately 1.44 in the wavelength range under
consideration \cite{malitson-1965-silica}.

In order to provide a resonant light reflection without forming the
diffraction orders, the dielectric bars are chosen to have the
resonant wavelength larger then the unit cell size. The size of the
dielectric bar cross-section is restricted by the following
condition $l\sqrt{\varepsilon}<\lambda/2$ where $\varepsilon$ is the
bar relative permittivity and $l$ is the cross section size; this
eliminates the transverse interference resonances inside bars.

The bar permittivity value is substantially limited by the
properties of dielectric materials suitable for the dielectric array
manufacture. The usual dielectrics extensively used in microchip
technologies have permittivities not exceeding 4. However, some
semiconductors can be used as materials for the array elements. The
use of semiconductors is promising because they have the
transparency windows in the visible and near-infrared wavelength
ranges. The permittivity of well-known suitable semiconductors has a
value from 11 to 18 and their dielectric loss tangent does not
exceed $10^{-3}$ within the transparency windows. For example, the
germanium transparency window extends from 1600~nm to 2000~nm and
its refractive index varies from 4.07 to 4.23 over this window
\cite{web-Germanium}.

Taking into account everything mentioned above, the size of the
square cross section of dielectric bars is chosen to be $l=195$~nm.

To solve the diffraction problem, the numerical method proposed in
\cite{khardikov-2008-pstd} is used. This method is based on both the
mapped PSTD method \cite{gao-2004-aam} and the transfer matrix
theory. For the simplicity the dispersion of dielectrics is not
taken into account in this paper. The relevancy of such approach is
due to two reasons. First, the dispersion of the chosen materials is
very weak in the wavelength range considered. Next this, dispersion
has no effect on the properties of the trapped-mode resonances.
However, the constituents dispersion may be taken into account in
the context of the method used.

\section{The trapped mode resonances in the arrays of dielectric bars}

Let us assume that silica fills up the all the half space $z>0$
below the array to simplify the analysis by excluding the
interference resonances in the substrate. As known, and shown by
simulation in \cite{khardikov-2008-pstd}, the finite thickness of
the substrate results in the interference fringes along the
wavelength dependences of the reflection and transmission
coefficients. In the actual structures approximately 0.5~mm thick,
these interference resonances are destroyed because of local
inhomogeneity of the substrate.

\subsection{The resonant properties of the arrays composed of metallic
bars}

For further comparison between the resonant properties of plasmonic
and entirely dielectric planar metamaterials, and to begin with,
let's briefly mention the reflection wavelength dependences of the
arrays of gold bars. The square unit cell of the structure
considered has $d_x=d_y=500$~nm. The longer and shorter bar lengths
are 450~nm and 400~nm respectively. The minimal distance between the
longer and the shorter bars is 100~nm. The array is placed on the
silica semi-infinite substrate. Since metals have a strong inherent
dispersion in the near infrared range, the gold dispersion has been
taken into account by using the complex-pole model and the method of
additional differential equation (see, for example
\cite{khardikov-2010-jop}).

\begin{figure}
\centering
\includegraphics[width=4.0in]{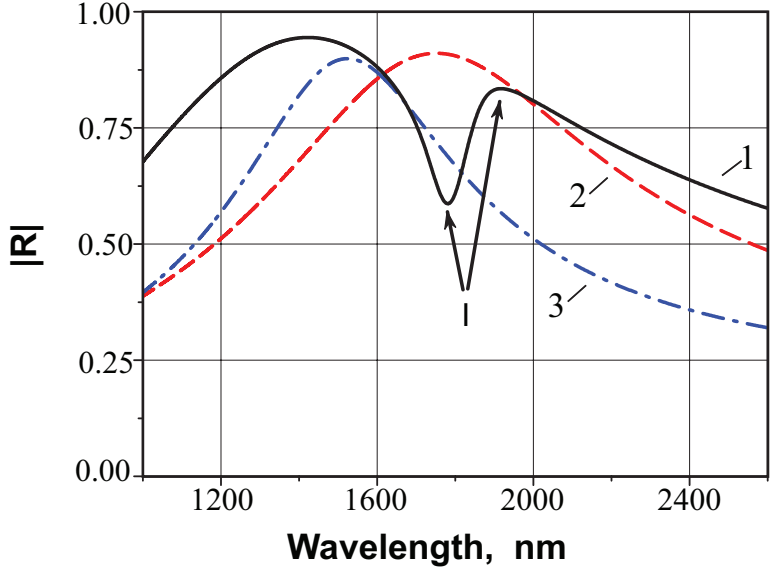}
\caption{Wavelength dependences of the magnitudes of the reflection
coefficients of the double-periodic structures composed of paired
gold bars ($h_1=450$~nm and $h_2=400$~nm, line 1) and single gold
bars whose length is 450~nm (line 2) or 400~nm (line 3). The sizes
of the periodic cell and the bar cross section are 500~nm and 50~nm,
respectively.} \label{fig2}
\end{figure}

The wavelength dependence of the reflection coefficient magnitude of
this structure is shown in Fig.~\ref{fig2} (line~1). Another two
lines present the wavelength dependences of the reflection
coefficient magnitudes of the arrays with a single gold bar in the
periodic cell. The line~2 corresponds to the array which consists of
the longer bars from the two-element array and the line~3 to the
array of the shorter ones.

The trapped mode resonance is marked by I in Fig.~\ref{fig2}. It has
the typical trough-and-peak Fano spectral profile
\cite{fano-61-spa}. As is well known \cite{khardikov-2010-jop,
zhang-2010-double_bars}, this resonance results from the excitation
of the anti-phased plasmon-polaritons in the adjacent metallic bars.
One important point is that the trapped mode resonance of the
two-element periodic structure is excited in the wavelength band
restricted by the wavelengths of the reflection resonances of the
one-element arrays or in a wavelength close to this band. It is
easily seen from the spectral dependences in Fig.~\ref{fig2} and
presented in \cite{zhang-2010-double_bars}.

\subsection{Identification of the reflection resonances of
a subwavelength dielectric array}

\begin{figure}
\centering
\includegraphics[width=4.0in]{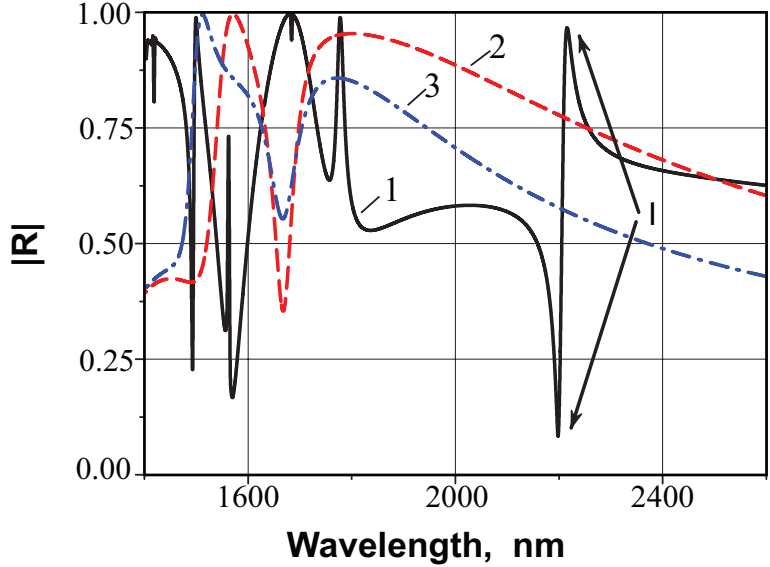}
\caption{The wavelength dependences of the reflection coefficients
magnitudes of the arrays composed of paired dielectric bars
($h_1=877$~nm and $h_2=780$~nm, line~1) and single dielectric bars
877~nm long (line~2) and 780~nm (line~3). The sizes of the periodic
cell and the bar cross section are 975~nm and 195~nm, respectively.
The refractive index of the bar dielectric is 5.5.} \label{fig3}
\end{figure}

First of all, in order to identify the kind of resonances, is
considered the normal reflection by arrays which are assumed to be
made of a hypothetic lossless dielectric with the large refractive
index 5.5. Such special choice of the refractive index enables to
observe at least two reflection resonances of one-element arrays of
dielectric bars in the subwavelength range. The wavelength
dependences of the reflection coefficients magnitudes are shown in
Fig.~\ref{fig3} for both the array of paired dielectric bars
(line~1) and the one-element arrays. The lines~2 and 3 correspond to
the array of the longer and the shorter dielectric bars,
respectively. There are two reflection resonances of the one-element
arrays in the considered wavelength range. As revealed by the
numerical analysis of the electric field distributions along the
periodic cell of arrays, they are the ordinary resonances of
dielectric bars. These resonances are excited under the condition
that the bar length is approximately equal to $\lambda_d/2$ or
$3\lambda_d/2$ where $\lambda_d$ is the wavelength in the dielectric
of bars.

If two bars of different length are combined in the periodic cell,
some additional reflection resonances are excited (see
Fig.~\ref{fig3}, line~1). Especial interest is stimulated by the
additional resonance in the long-wave part of the considered range.
It is a typical Fano-shape sharp resonance with a specific
wavelength dependence of the reflection coefficient rolled over deep
to peak.

\begin{figure}
\centering
\includegraphics[width=4.0in]{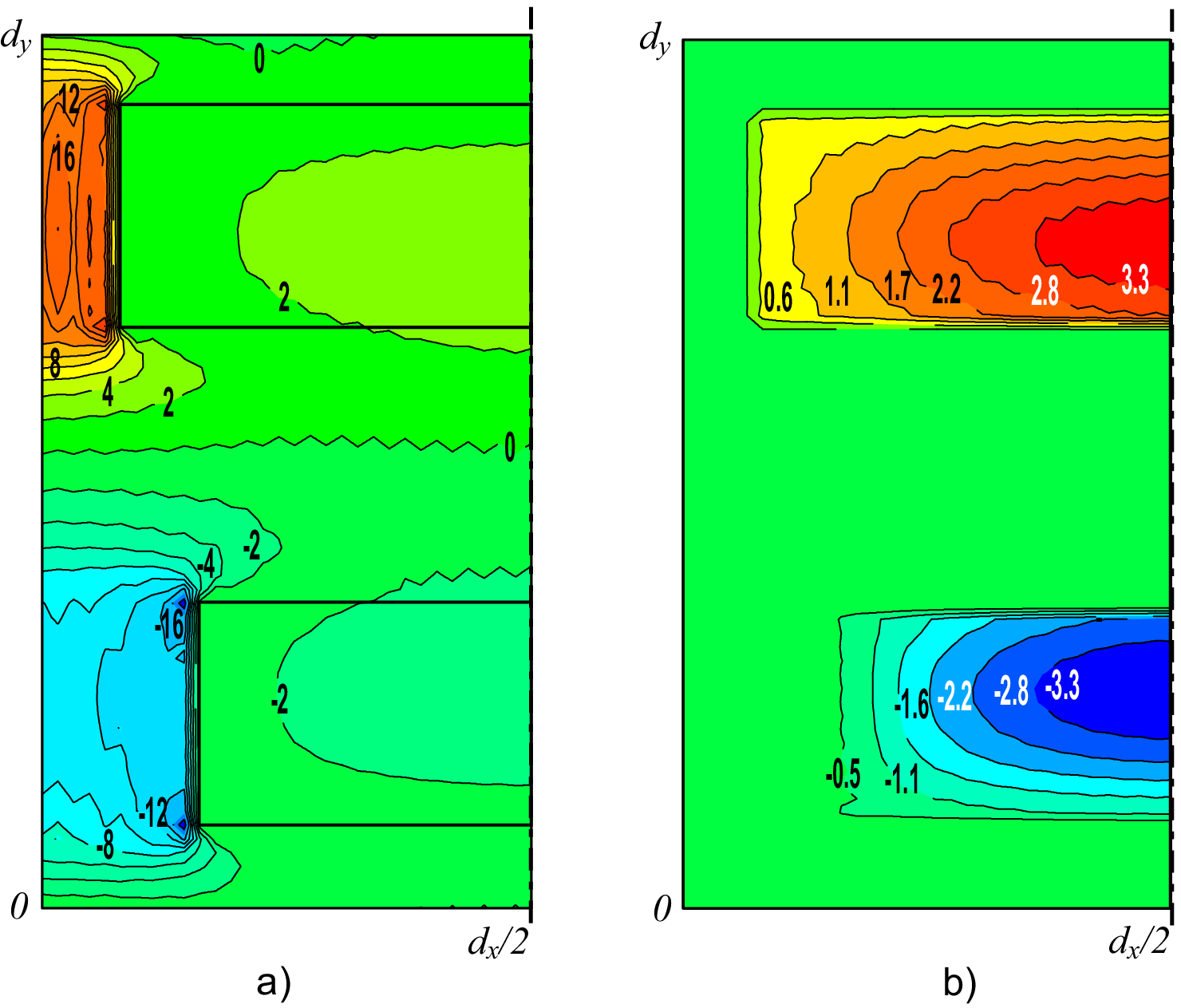}
\caption{The distribution of the electric field $x$-component within
the periodic cell of the two-element dielectric array in the plane
$z=l/2$. The field distribution is presented for the longest
resonant wavelength 2208~nm (see Fig.~\ref{fig3}). Since the
electric field values essentially differ inside and outside the bars
the field distribution is plotted, for convenience, in the whole of
the periodic cell (a), and only inside the bars (b) at different
scales.} \label{fig4}
\end{figure}

The electric field distributions within the periodic cell have been
plotted and studied over for all of the additional resonances, to
clarify their nature. One of such distributions relating to the
plane $z=l/2$ is shown in Fig.~\ref{fig4} at the wavelength 2208~nm
that corresponds to the additional resonance with the maximum
wavelength. As the field has a symmetric distribution, it is
presented only within a half of the cell.

One can see that the both dielectric bars behave as a
half-wavelength dielectric resonators, i.e. the maximum of the
electric field within the bar locates close to its center and the
field decreases to the bar facets (Fig.~\ref{fig4},b). The electric
field maxima observed outside the bar at its ends result from large
enough difference between the permittivities inside and outside the
bar. Actually, the normal components of the electric field induction
satisfy the continuity condition $E_{x-}=\varepsilon E_{x+}$, where
$E_{x-}$ and $E_{x+}$ are the electric fields in free space and
inside the bar, respectively. Thus the electric field in free space
between the bars has the value $\varepsilon$ times grater than the
field inside the bars. In our case this coefficient is
$\varepsilon=30.25$. The antiphased field distribution evidences
that the  studied resonance is a trapped mode one. Notice that the
$E_y$ field component has its maximum value of the same order as the
maximum of the $E_x$ and the maximum of $E_z$ one is over then 100
times less.

Besides the enhanced Q-factor, the main distinctive feature of the
trapped-mode resonance of the two-element dielectric array is a
great red shift of its wavelength relative the resonant wavelengths
of the corresponding one-element arrays (see Fig.~\ref{fig3}). Thus
the coupling between the dielectric bars of the two-element periodic
array results in an extremely large increase of the resonant
wavelength as opposed to the coupling in the array of metallic bars
(see Fig.~\ref{fig2}). This property of the entirely dielectric
trapped-mode arrays is quite important in view of possible
applications in the field of infrared metamaterials. First, the
ratio of the array pitch to wavelength may be decreased to design
more homogeneous metamaterials. Second, the increase of the resonant
wavelength results in an enhancement of the confinement of the field
intensity due to a decrease of radiation losses. It is equally
important to designing of nonlinear and active artificial media.
Third, this property of dielectric arrays gives us a way to design
double-periodic structures with the trapped mode resonance using
materials of relatively small refractive index; for example, made of
semiconductors in the wavelength range of their transparency
windows.

Finally, the research of the field distributions concerning all the
rest resonances of the studied over two-element dielectric array
(see the spectral dependence in Fig.~\ref{fig3}, line~1) results in
the conclusion that they are the ordinary dimensional resonances.

\subsection{The trapped mode resonance of a germanium-bar array
within the transparency window bandwidth}

\begin{figure}
\centering
\includegraphics[width=4.0in]{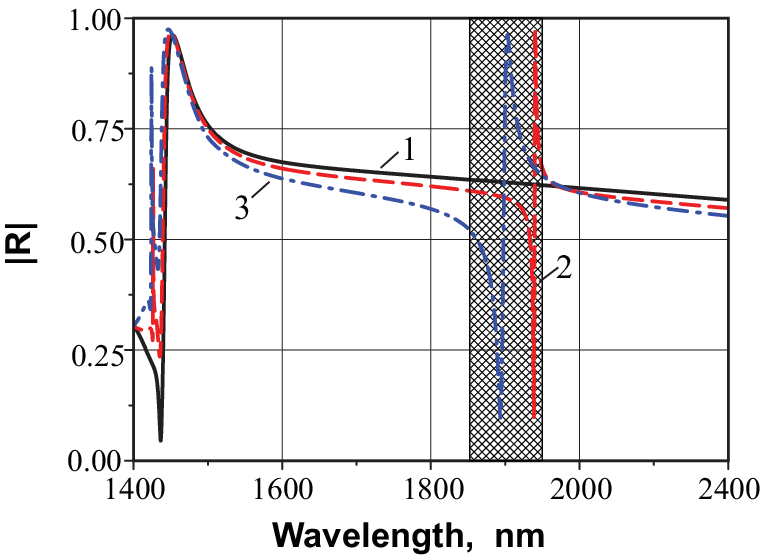}
\caption{The wavelength dependences of the reflection coefficients
of \textbf{the} arrays of paired germanium bars. The structure sizes
are $d_x=d_y=975$~nm, $l=195$~nm, $h_1=877$~nm. Lines 1, 2 and 3
correspond to $h_2=877$~nm, $h_2=838$~nm and $h_2=780$~nm
respectively. It is assumed that the refractive index of germanium
is 4.12, and the dissipative losses are negligibly small.}
\label{fig5}
\end{figure}

Now let us study some more realistic array of the germanium bars
with the sizes mentioned above. The refractive index of germanium is
assumed to be equal to 4.12. Such value of the germanium refractive
index corresponds to the wavelength range from 1850 to 1950~nm. This
range is a shaded one in Fig.~\ref{fig5} where the wavelength
dependence of the reflection coefficient of the array is presented.
The trapped mode resonance is observed in this range.

Fig.~\ref{fig5} illustrates the effect of the array asymmetry as a
consequence of the influence of the difference of bar lengths on the
wavelength and the Q-factor of the trapped mode resonance. It is
evident that the trapped mode resonance cannot be excited in a
symmetrical structure with equal lengths of both bars by the
normally incident plane wave because of the zeroth coupling of
asymmetrical mode of bars and plane wave in the free space and
substrate (see line~1 in Fig.~\ref{fig5}). A decrease of the
asymmetry degree characterized by the value $h_1/h_2$ results in an
increase of the red shift of the trapped mode resonance and a
decrease of the wavelength difference between the reflection deep
and peak, i.e. in the trapped mode resonance Q-factor growth.

To estimate the trapped mode resonance Q-factor consider the
expression $Q=\lambda_1\lambda_2/(2\lambda_0(\lambda_2-\lambda_1))$,
where $\lambda_1$ and $\lambda_2$ are the wavelengths of the deep
and peak of the reflection coefficient, respectively, and
$\lambda_0$ is the wavelength corresponding to the reflection
coefficient value $(|R(\lambda_1)|+|R(\lambda_2)|)/2$. The Q-factors
of the trapped mode resonances of the germanium bars structures are
203 and 1080 for $h_1=877$~nm, $h_2=780$~nm and $h_1=877$~nm,
$h_2=838$~nm, respectively (see the lines 2 and 3 in Fig.~\ref{fig5}
for comparison). Note that for the structure of the dielectric bars
with the refractive index 5.5 (see Fig.~\ref{fig3}, line 1), the
Q-factor of the trapped mode resonance is 127.

\begin{figure}
\centering
\includegraphics[width=4.0in]{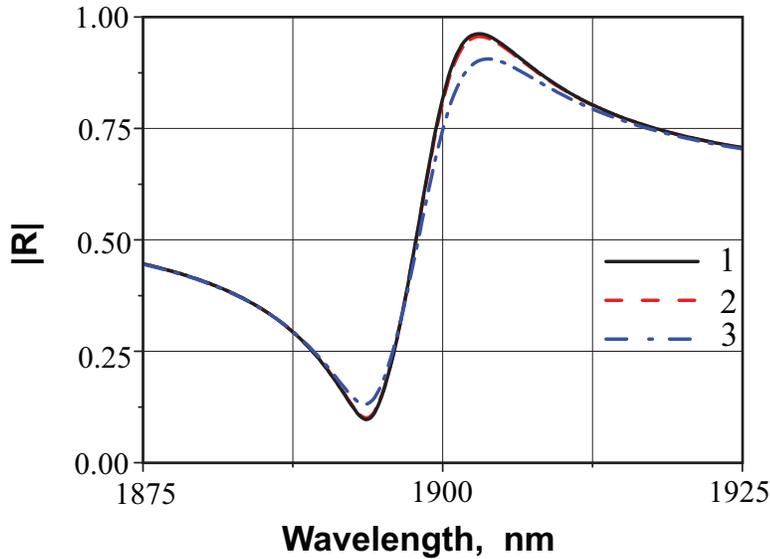}
\caption{The wavelength dependences of the reflection coefficient of
the array composed of paired germanium bars with dissipation. The
structure sizes are $d_x=d_y=975$~nm, $l=195$~nm, $h_1=877$~nm and
$h_2=780$~nm. It is assumed that the refractive index of germanium
is 4.12 and the dielectric loss tangent has the values
$\tan(\delta)=0$ (line 1), $\tan(\delta)=10^{-4}$ (line 2), and
$\tan(\delta)=10^{-3}$ (line 3).} \label{fig6}
\end{figure}

The semiconductors dissipative losses will certainly effect the
Q-factor of the trapped mode resonance. However, in the transparency
window the dielectric loss tangent of germanium does not exceed
$10^{-3}$ (this bandwidth is marked as shaded in Fig.~\ref{fig5}).
Over this band, one can see only negligible variations of the
reflection magnitude and a very small widening of the trapped mode
resonance which reveals a decrease of the Q-factor (see
Fig.~\ref{fig6}) with an increase of the dielectric loss tangent.

The field of high intensity confined in the periodic array under the
trapped mode resonance may effectively interact with a substrate of
an active or nonlinear material. In the first case, one can observe
amplification of the reflected or the transmitted light. In the case
of a nonlinear substrate, a certain light with light controlling can
be achieved in very thin structures.

\begin{figure}
\centering
\includegraphics[width=4.0in]{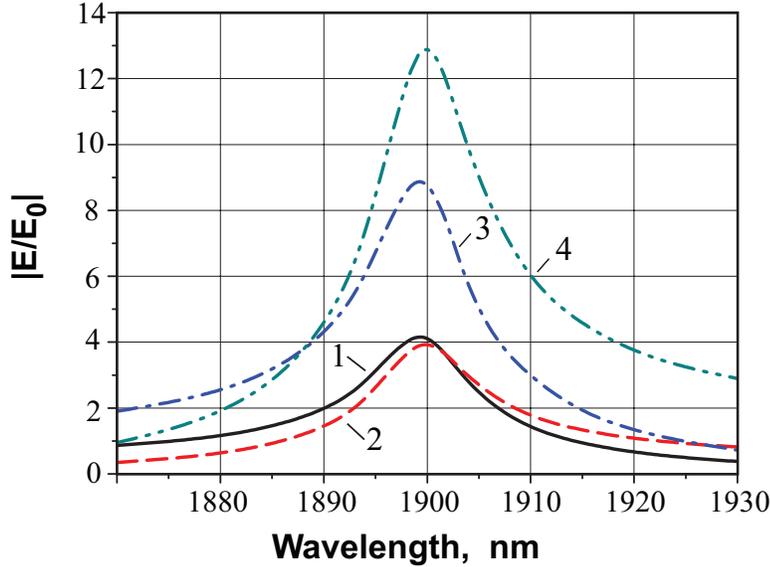}
\caption{Wavelength dependences of ratio $|E|/|E_0|$ for some points
within two-element array of germanium bars. The structure sizes are
$d_x=d_y=975$~nm, $l=195$~nm, $h_1=877$~nm and $h_2=780$~nm.
Refractive index of germanium is assumed to be 4.12. Lines 1 and 2
correspond to the center of the shorter and the longer bars,
respectively. Lines 3 and 4 correspond to the center of air gaps
between two adjacent shorter and longer bars, respectively.}
\label{fig7}
\end{figure}

The wavelength dependences of the ratio $|E|/|E_0|$ are presented in
Fig.~\ref{fig7}. Here $|E|$ is the maximum of the absolute value of
the electric field in one or another chosen point within the array
and $|E_0|$ is the maximum of the absolute value of the incident
wave electric field in free space. The wavelength dependences of
$|E|/|E_0|$ are shown for four different points. Lines 1 and 2
correspond to the centers of the shorter and the longer dielectric
bars, respectively. Near those points the electric field reaches its
maximum level inside the bars. Lines 3 and 4 correspond to the
center of a gap between two adjacent shorter and longer bars,
respectively. It should be noticed that the electric field magnitude
reaches its minima along the both mentioned gaps in these points
(see Fig.~\ref{fig4},a). The magnitude of electric field enlarges
over 28 times in the center of the gap between the bars despite the
fact that the electric field has its minimum in these points. The
maximum of the electric field magnitude in array corresponds to the
wavelength 1899~nm. This wavelength is approximately equal to the
central wavelength of the trapped mode resonance
($(\lambda_1+\lambda_2)/2=1898$~nm).

The Q-factor increasing results in a rise of the electric field
intensity in the structure. For example, the maximal ratio
$|E|/|E_0|$ is 2.3 times greater in the case of the germanium array
with $h_1=877$~nm and $h_2=838$~nm than with the array with
$h_1=877$~nm and $h_2=780$~nm.

\section{Conclusions}

The problem of the normal reflection of the near-infrared radiation
by a planar array with a subwavelength square translation cell
composed of two dielectric bars of different lengths has been
solved. For the fist time the existence of the high-Q trapped-mode
resonance has been found out in these low-loss entirely dielectric
structures. A coupling between the adjacent bars induces the field
enhancement in the surrounding media; which can enforce the
phenomena like luminescence, nonlinear scattering, absorption, and
lasing. The spectral response of this novel planar metamaterial
closely resembles the EIT phenomenon in atomic systems. In contrast
with plasmonic arrays, the trapped-mode resonance excited in an
entirely dielectric structure demonstrates a giant red shift
relative to the wavelength of the ordinary resonance of array with
only one dielectric bar per unit cell. This remarkable property of
the dielectric trapped-mode arrays enables to design the highly
desirable deep-subwavelength low-loss planar metamaterials in near
infrared range.

\ack

This work was supported by the Ukrainian State
Foundation for Basic Research, Project F40.2/037.


\section*{References}


\bibliographystyle{vancouver}

\bibliography{GiantRedShift_5}


\end{document}